\documentclass[aps,pre,twocolumn,groupedaddress,showpacs]{revtex4}
\usepackage{graphicx}
\usepackage{amssymb}

\begin{document}


\title{Jamming and growth of dynamical heterogeneities versus depth for granular heap flow}


\author{H. Katsuragi,\footnote{Current address: Department of Applied Science for Electronics and Materials, Kyushu University, Fukuoka 816-8580, Japan}  A. R. Abate,\footnote{Current address:  Department of Physics, Harvard University, Cambridge, MA 02138, USA} and D. J. Durian}
\affiliation{Department of Physics and Astronomy, University of Pennsylvania, Philadelphia, PA 19104-6396, USA}


\date{\today}

\begin{abstract}
We report on grain dynamics versus depth for steady-state gravity-driven flow of grains along a heap formed between two parallel sidewalls.  Near the surface the flow is steady and fast, while far below there is no flow whatsoever; therefore, a jamming transition occurs where depth is an effective control parameter for setting the distance from the transition.  As seen previously, the time-averaged velocity profile along the sidewall exhibits a nearly exponential decay vs depth.  Using speckle-visiblility spectroscopy (SVS), we find that velocity fluctuations grow relative to the average on approach to jamming.  Introducing an image-based order parameter and the variance in its temporal decay at a given depth, we also characterize the spatially heterogeneous nature of the dynamics and find that it increases on approach to jamming.  In particular, the important time and length scales for dynamical heterogeneities are found to grow almost exponentially as a function of depth.  These dynamical changes occur without noticeable change in structure, and are compared to behavior found by experiment and simulation elsewhere in order to test the universality of the jamming concept.  In particular we find that the size of the heterogeneities scales as the inertia number raised to the $-1/3$ power, in agreement with two recent simulations.
\end{abstract}


\maketitle



``Jamming'' is defined to occur when a disordered system develops a yield stress, or alternatively, as a practical matter, when it develops a stress relaxation time that exceeds a reasonable experimental time scale \cite{OHernPRE03}.   Granular materials are disordered systems that readily jam, for example as a static heap that maintains its slope indefinitely against gravity  \cite{Jaeger1996, Duran, Midi2001}.   The general behavior of granular systems remains a topic of fundamental interest in part because of possible similarities with dense colloids, supercooled liquids, and soft glassy materials, in their approach to jamming \cite{CatesPRL98, LiuNagelBook}.  For example the microscopic dynamics generally exhibit spatiotemporal heterogeneities that are expected to increase in size as the relaxation times become longer \cite{SillescuSHDinSCL, EdigerSHDinSCL, GlotzerJNCS00, CipellettiRamos05}.  Such heterogeneities have been directly visualized in dense colloidal suspensions \cite{ MarcusPRE1999ColloidalStrings, KegelScience2000ColloidalStrings, WeeksWeitzRelaxation} and in granular materials subjected to shear \cite{PouliquenPRL03, DauchotPRLSubDiff2005, DauchotSHDPRL2005}.  They have also been visualized for large grains in a nearly fluidizing upflow of air \cite{KeysAbateNP07}, and have been found to increase on approach to jamming with a dynamic correlation length exponent  \cite{AdamPRE07, AbatePRL08} in agreement with diverging length scales observed in simulations \cite{OHernPRE03, ReicchardtPRL05, OlssonTeitelPRL07}.

To explore and possibly extend the universality of the jamming concept, it is crucial to broaden the range not just of systems but also of different control parameters that can be varied to adjust the proximity to jamming.   Toward this end, in this paper, we consider the flow induced by the steady addition of grains to the top of heap confined between parallel plates.  Whereas earlier with this setup we studied behavior by varying the rate of flow \cite{PierrePRL00, AdamHiroPRE07}, now we fix the flow rate and study behavior as a function of depth below the surface.  In this regime of continuous flow, the velocity profile along the sidewall decays nearly exponentially with depth \cite{PierrePRL00}; this observation has been reproduced and explored in detail \cite{Khakhar2001, Komatsu2001, Andreotti01, Jop2005, Djaoui05, Richard08}.  In effect the grains are strongly driven and unjammed near the surface but become progressively jammed as a function of depth.  To quantify this jamming transition, we measure the coarse-grained surface velocity profiles with great dynamic range and we also implement two novel techniques to characterize grain-scale behavior.  In particular we use Speckle-Visibility Spectroscopy (SVS) \cite{SVSreview} to measure the time-resolved relative speed of grains, and we introduce a video-based dynamic order parameter and associated four-point susceptibility to measure dynamical heterogeneities, all as a systematic function of depth.  We find that fluctuations and heterogeneities increase with depth, and we compare with similar behavior reported for prior approaches to jamming in both experiment and simulation.


\section{Experimental system}

\begin{figure}
\includegraphics[width=3.00in]{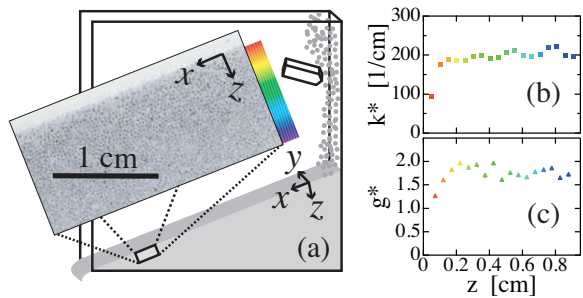}
\caption{(Color online). (a) Sketch of the experimental apparatus and coordinate system, plus callout of one raw video frame with inverted grayscale.  The color code is used in subsequent figures to indicate depth.  (b) Characteristic wave number $k^*$ in $x$ direction, vs depth $z$, from the power spectrum of video images.  (c) Peak value of the pair correlation function $g(x)$, vs depth $z$, from particle-tracking analysis of video images.} \label{akg}
\end{figure}

The flow cell and granular medium are identical to that of Refs.~\cite{PierrePRL00, AdamHiroPRE07}, in which avalanches were studied by slow steady addition of $d=\{0.25-0.35\}$~mm diameter glass beads at the top of a heap formed between parallel static-dissipating Lucite plates.  The width of the gap between the plates is $w=0.95$~cm, and is closed along the bottom and one vertical side; the inner area is square $28.5\times28.5$~cm$^2$.  The static granular heap has angle of repose $\theta_r=24^\circ$ and bulk mass density $\rho=1.5$~g/cm$^3$, which corresponds to spheres of mass density 2.4~g/cm$^3$ arranged with packing fraction 0.64.

For all experiments, grains are fed onto the top of the heap at the back of the box at a fast steady rate of $Q=2.5$~g/s.  This is accomplished using a large funnel with a fine control valve, and an aluminum bar sandwiched between the plates to break the free-fall of grains and deposit them gently onto the heap.  Long-duration runs are accomplished by collecting the discharge and pouring it back into the funnel.  Note that the choice of flow rate onto the top of the heap is nearly ten times greater than the critical rate $Q_c=0.36$~g/s~\cite{PierrePRL00, AdamHiroPRE07}, below which the discharge out the bottom is intermittent and accomplished by a series of discrete avalanches.  Therefore the measurements below are all for a regime in which the flow is smooth, continuous, and independent of time.  Over an extended distance, several channel widths in from the entrance and exit, the flow field appears parallel to the top free surface oriented at $\theta=26^\circ$, two degrees higher than the repose angle.  Care is taken that all measurements are conducted in this fully developed region, such as illustrated in Fig.~\ref{akg} where the example image is acquired about 5~cm from the exit.

Characteristic scales may be roughly estimated by assuming that the average velocity decreases to zero over a depth set by the channel width.  This gives a characteristic flow speed of $Q/(\rho w^2)=2$~cm/s, which is nearly equal to the projection of grain-size free-fall speed along the surface, $\sqrt{g d}\sin\theta_r$.  The characteristic strain rate is given by speed divided by channel width as $\dot\gamma=2$~s$^{-1}$; therefore, the time 0.5~s to accumulate a strain of one is much longer than the grain collision time, which must be shorter than $\sqrt{d/g}=5.5$~ms.  This means that the grains either experience many rapid collisions within a long-lived set of neighbors, reminiscent of caging in a glassy system, or else they flow subject to enduring contacts.  Taking pressure as $P=\rho g w$, the characteristic ``inertia number'' is $I=\dot\gamma d / \sqrt{P/\rho}=0.002$; this means that inertial stresses are small compared to confining stresses~\cite{Midi2001}.


\section{Structure}

Using a high-speed Phantom video camera, $960\times480$ pixels and 8~bit gray scale, we begin by acquiring a sequence of still images as shown for example in the callout of Fig.~\ref{akg}.  For this, the camera is mounted about 30~cm away from the side of the sample, oriented at the same angle as the heap, and fitted with a macro lens.  The magnification is 43.6~pixels/mm, or about 13 pixels per grain diameter.  Individual grains along the wall are thus readily visible, and the intensity profile across grains is a series of smooth peaks and valleys (not step functions); therefore,
their coordinates may be found by standard particle-tracking methods.  By counting the number of particles in a narrow 22-pixel slice for different depths, we find that the number density rises from zero to a constant, presumably random-close packing, by a depth of only a couple grains.  Two other measures of instantaneous packing structure exhibit similar behavior.  The first, Fig.~\ref{akg}(b), is for the location $k^*$ of the peak in the power spectra for image slices at different fixed depths.  This rises to a constant exactly as for the number density profile.  The second, in Fig.~\ref{akg}(c), is for the value $g^*$ of the first peak in the pair correlation function, computed from the grain coordinates found in 22-pixel image slices at different fixed depths.   This, too, rises to a constant but somewhat more gradually than $k^*$ and number density data.  By all three measures, the influence of the top free surface on the packing structure of the flowing grains is limited to a shallow layer of only a few grains, much less than the width of the channel.  This result agrees with numerical simulations for flow on an inclined plane~\cite{Silbert01,Mitarai05}.  Exactly as for the jamming and glass transitions, the dramatic changes next discussed for dynamics vs depth into the heap are accomplished without significant change in structure.


\section{Velocity Profiles}

\begin{figure}
\includegraphics[width=3.00in]{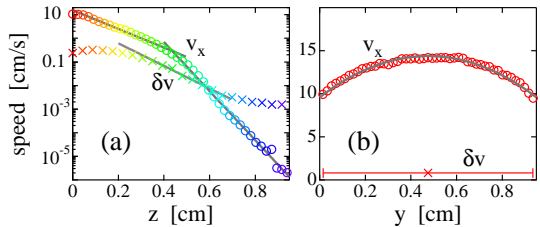}
\caption{(Color online).  (a) Mean speed $v_x$ in the flow direction, from video data, and fluctuation speed $\delta v$, from speckle-visibility spectroscopy data, vs depth $z$ along the sidewall.  (b) Mean and fluctuation speeds vs distance $y$ along the top free surface at $z=0$.  The depth is redundantly indicated using the color code shown in Fig.~\protect{\ref{akg}}.  The gray lines in (a) represents exponentials and the and gray curve in (b) is a parabola.}
\label{v}
\end{figure}

The same video set-up and particle-tracking methods as above can be used to measure the average speed $v_x(z)$ of flow in the $x$-direction along the wall as a function of depth $z$.   For this the images are sliced into regions 10-pixels wide, and the average velocity at a given depth is found by the grain displacement in the $x$ direction between successive frames, multiplied by frame rate, averaged over all particles in the slice, and averaged for 1000 frames.   To measure faster speeds near the free surface as well as slower speeds down deep, this is done for a wide range of frame rates: \{3900,  2048, 1024, 512, 256, 128, 64, 32, 16, 8, 4, 2, 1, 0.5, 0.25\} per second.  The results for all frame rates are averaged together with weighting given by the respective statistical uncertainties.  The weight of a contribution is highest when the average displacement is about $1/10$ grain per frame.  A similar procedure is also employed to measure the average speed $v_x(y)$ of the flow at the free surface as a function of the distance $y$ away from the front plate.  For this we use 1000 frames, $576\times576$ pixels, taken at a rate of 4000 per second.

The mean velocities along the wall and across the top free surface are displayed in Fig.~\ref{v}.  The flow profile along the wall, Fig.~\ref{v}(a), begins at 10~cm/s and decays by a factor of nearly $10^6$ over a distance of about 1~cm.  The shape appears to cross over between two exponentials at about $0.4$~cm.  Exponential decay has been reported for both heaps \cite{PierrePRL00, Khakhar2001, Komatsu2001, Andreotti01, Jop2005, Djaoui05, Richard08} and rotating drums~\cite{duPont2005}; similar profiles have been observed in Couette cells \cite{Howell99, Meuth00, Bocquet02}.  The flow profile across the top surface, Fig.~\ref{v}(b), also begins at 10~cm/s but increases much less dramatically to a maximum of 15~cm/s at the center; the shape is nearly parabolic.  Assuming such a parabolic increase away from the wall at all depths, we integrate the granular flux over the entire cross section and arrive at a mass discharge rate within ten percent of the known value $Q=2.5$~g/s.  Based on this check we suppose that the behavior along the sidewall is at least somewhat representative of that in the interior.  The influence of the side wall on bulk behavior was studied in Ref.~\cite{Jop2005}.


\section{Velocity Fluctuations}

\begin{figure}
\includegraphics[width=3.00in]{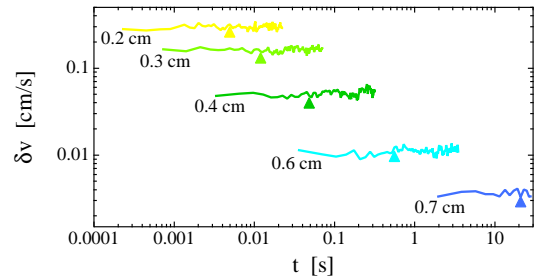}
\caption{(Color online). Fluctuation speed $\delta v$ vs time for several depths, $z$, as labeled.  The curves are decorated with a solid triangle at the time $r/v_x$ needed to advance with the average flow by one grain radius.}
\label{dv_t}
\end{figure}

\begin{figure}
\includegraphics[width=3.00in]{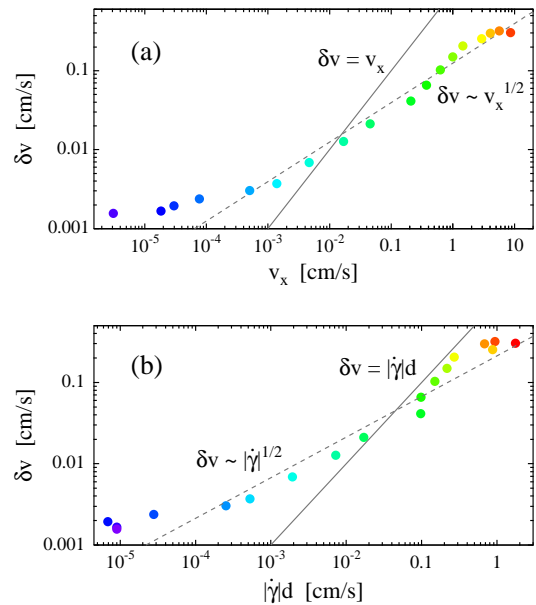}
\caption{(Color online).  Fluctuation speed $\delta v$ vs (a) average flow speed $v_x$ and vs (b) strain rate times grain diameter.  Each point corresponds to a different depth, as indicated by the same color code as in the previous figures.  In each plot the solid line represents $y=x$; in (a) the dashed line represents the best fit to a square-root power law.}
\label{dv_gdvx}
\end{figure}

Next we measure the size $\delta v$ of microscopic velocity fluctuations by use of Speckle-Visibility Spectroscopy (SVS)~\cite{SVSreview}.  This is a time-resolved dynamic light scattering method, which we have previously used to measure $\delta v$ vs time for intermittent avalanches on the top of a heap~\cite{AdamHiroPRE07}.   One advantage of SVS over particle tracking for measurement of fluctuations is that the incident photons provide an ensemble average over an interior region several grains from the boundary. Another advantage is that it probes wavelength-scale motion much smaller than the grain size; this is particularly helpful on approach to jamming where the persistence length for ballistic motion becomes so small that particle tracking would return an average speed contaminated by multiple direction changes due to neighbor collisions.  Here we use the same setup as before~\cite{AdamHiroPRE07}: illumination by Nd-YAG laser with wavelength $\lambda = 532$~nm and maximum power 4~W; detection by 8-bit linescan camera with $2\times 1024$ pixels and 58~kHz frame rate.  The main difference is that the incident light is spread out with a cylindrical lens to form a line about 5~cm long and 1~mm wide at variable depth parallel to the top free surface of the heap.  The depth-resolution is further smeared beyond 1~mm because incident photons migrate within the sample, primarily over a region set by several grain diameters.  The camera is placed about 10~cm away, so that the speckle size is closely matched to the pixel size.  As the grains move {\it relative to one another}, the speckle pattern changes and hence each image blurs out to an extent that depends on the duration $T$ of the exposure.  This effect is quantified by the variance $V(T)$ in detected intensity levels, which we compute for a sequence of synthetic exposure durations by addition of a given number of successive frames of video data.  Here, where the laser light is diffusely backscattered from ballistically moving particles, the spectrum is Lorentzian with some linewidth $\Gamma$.  The theory of SVS \cite{SVSreview} gives the variance ratio for exposures of $2T$ and $T$ duration as
\begin{equation}
    { V(2T) \over V(T) } = {e^{-4x}-1+4x \over 4(e^{-2x}-1+2x) }
\label{SVS}
\end{equation}
where $x \equiv \Gamma T \approx (4\pi\delta v/\lambda)T$.  The quantity $\delta v$ represents an ensemble average over the scattering volume of the characteristic speed for the relative motion of grains separated by one photon transport mean free path, $l^*$.  In general this includes contributions from randomly-directed fluctuations $\delta v_{rand}$ as well as from shear.   For pure shear deformation, with no fluctuations, at absolute rate $|\dot\gamma|$, the time scale $\tau_S$ in Eq.~(7) of Ref.~\cite{Wu90} gives the expectation for $\delta v$ as $|\dot\gamma| l^*/\sqrt{5}$; this is the same relationship noted previously for Diffusing-Wave Spectroscopy (DWS)  \cite{PierrePRL00, Richard08}.  Since $l^*$ is a few grain diameters \cite{LeutzRicka96, Menon1997, PierrePRL00, Crassous07, Djaoui05}, the relative speed expectation reduces to approximately $ |\dot\gamma| d$ for pure shear.  If the grains additionally experience random ballistic velocity fluctuations around the mean shear flow, then the relative speed deduced from Eq.~(\ref{SVS}) will be commensurately larger than this lower bound as $\delta v = \delta v_{rand}+|\dot \gamma | d$.

Example data for the instantaneous characteristic relative speed $\delta v$ vs time are displayed in Fig.~\ref{dv_t} for five different depths.  The relative speeds decrease with increasing depth, and appear to be constant in time.  We anticipated, but did not find, temporal fluctuations in $\delta v$ that increase with depth.  Such fluctuations could be present, but remain unobserved due to the spatial averaging in proportion to illumination area and the resulting photon sampling volume.  In either case, we average each such time sequence to arrive at a single value for the mean relative speed at a given depth.  The results are plotted in Figs.~\ref{v} along with the profiles of mean speed.  At the top free surface, Figs.~\ref{v}(b), the value of $\delta v$ is much less than the mean speed.  Thus the instantaneous grain velocities point primarily in the flow direction, with only slight fluctuations around the average.  The same is true along the wall but only near the top, Figs.~\ref{v}(a). In particular $\delta v$ decays with depth less rapidly than $v_x$, so that below 0.6~cm the relative grain speed becomes greater than the mean flow speed.  Below this depth the grains don't merely flow but must jostle more and more with their neighbors.  This can be thought as similar to the onset of caging effects on approach to jamming.

To examine the relationship between fluctuation and flow, we display $\delta v$ vs $v_x$ on logarithmic axes in Fig.~\ref{dv_gdvx}(a).  This is a parametric plot, in that each point represents a single depth.  For the upper two decades in $\delta v$, and the corresponding upper four decades in $v_x$, a square-root relationship $\delta v \sim \sqrt{v_x}$ holds quite well as shown in Fig.~\ref{dv_gdvx}(a) by the dashed line of slope $1/2$.  Since $\delta v$ and $v_x$ are not proportionate, there is no quasistatic limit where the nature of the dynamics merely slows down without alteration of character.  Similar behavior was measured for continuous flow in a hopper \cite{Menon1997} and also at the free surface of a heap \cite{AdamHiroPRE07}, for both of which the flow and fluctuation speeds were decreased by a reduction in discharge rate.  The same behavior, $\delta v \sim \sqrt{v_x}$, is also implied by the simulation result $\delta v^2 \propto \dot\gamma$ \cite{HatanoJPSJ08}, since here the velocity profiles are nearly exponential; this is seen directly in Fig.~\ref{dv_gdvx}(b) where $\delta v$ data are plotted vs strain rate.   In all cases the relative rise of $\delta v$ over the flow speed for greater depth signals increased jostling and dissipation at decreased driving rate.  For the hopper this causes clogging; at the top of a driven heap this causes a transition from steady flow to intermittent avalanches; here along the sidewall for steady heap flow this causes jamming at great enough depth.

The separate contributions of fluctuation and shear to the SVS signal may be judged by the plot of $\delta v$ results vs $|\dot\gamma|d$ in Fig.~\ref{dv_gdvx}(b).  Since the flow profile is roughly exponential, this plot appears similar to $\delta v$ vs $v_x$ in part (a) of this figure.  However now we may compare directly with the prediction $\delta v \approx |\dot\gamma|d$ for pure shear.  Indeed for the rapid flow near the surface, the data roughly follow this form.  But deeper below the surface, the $\delta v$ data rise above $|\dot\gamma|d$.  This is an alternative way of demonstrating that the grains jostle around the mean flow to an extent that is small near the surface but that increases systematically vs depth on approach to jamming.


\section{Dynamical Heterogeneities}

\begin{figure}
\includegraphics[width=3.00in]{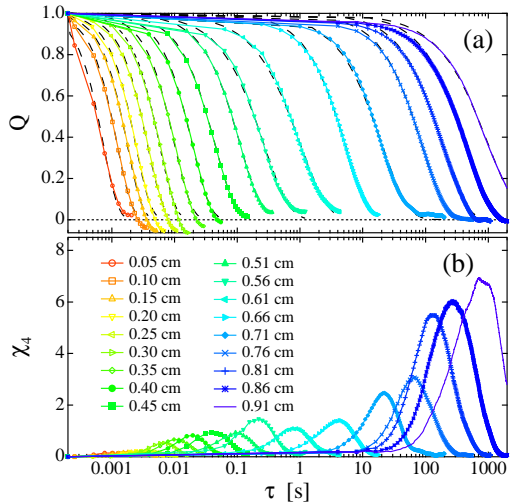}
\caption{(Color online). (a) The average overlap order parameter $Q$, and (b) the corresponding susceptibility $\chi_4$ as functions of lag time $\tau$, for several depths as indicated by color and legend.   Dashed curves in (a) represent fits to a compressed exponential, $Q(\tau)=\exp[-(\tau/\tau_0)^\beta]$, where both $\tau_0$ and $\beta$ are adjusted. }
\label{QX4_t}
\end{figure}

\begin{figure}
\includegraphics[width=3.00in]{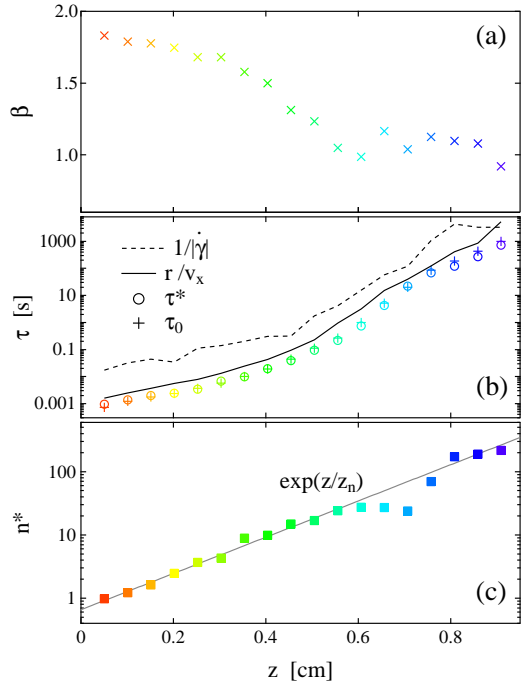}
\caption{(Color online). (a) Exponent $\beta$, and (b) decay time scale $\tau_0$, for the compressed exponential fits in Fig.~\protect{\ref{QX4_t}}(a), versus depth $z$.  The time $r/v_x$ to flow one bead radius, the time $\tau^*$ at which $\chi_4$ peaks, and $1/|\dot \gamma|$, are also included in (b) as labeled.  (c) The number $n^*$ of fast-moving grains in a dynamical heterogeneity versus depth.}
\label{btn_z}
\end{figure}

The transition to jamming in glasses, colloids, and granular media appears to be universally accompanied by the growth of dynamical heterogeneities.  Grain-scale motion is not possible unless there is cooperation with neighbors, for example in the form of string-like swirls that come and go; the closer to jamming, the more infrequent the swirls and the larger the number of grains involved.  A well-developed approach for characterizing such fluctuating correlated dynamics begins with an ensemble-averaged overlap order parameter $Q_t(\tau)$ vs delay time $\tau$ and its average $Q(\tau)=\langle Q_t(\tau)\rangle$ over all start times $t$; see the review in Ref.~\cite{Glotzer03}.  For a finite system of $N$ grains, spatially heterogeneous dynamics manifest as fluctuations in the instantaneous number of fast-moving regions and hence as variability in the decay of $Q_t(\tau)$ for different start times.  This is captured by the the normalized variance $\chi_4(\tau)=N[\langle Q_t^2(\tau) \rangle - \langle Q_t(\tau) \rangle^2]$, which is a four-point susceptibility that depends on the time interval $\tau$ and a spatial length usually comparable to the grain size.  Note that this definition renders $\chi_4(\tau)$ independent of $N$ since the variance for counting statistics decreases as $1/N$.  An explicit counting statistics argument for the average and variance of the number of fast-moving regions \cite{AdamPRE07} then gives the number $n^*$ of grains in a dynamical heterogeneity as
\begin{equation}
    n^* = { {\chi_4}^* \over (Q_1-Q_0)(Q_1 - Q^*) }
\label{nstar}
\end{equation}
where ${\chi_4}^*$ and $Q^*$ are the respective values where $\chi_4(\tau)$ is maximum, and where $Q_1$ and $Q_0$ are the average contributions at this time coming from the slow and fast regions respectively.  On approach to jamming, the values of ${\chi_4}^*$ and $n^*$ increase because greater cooperativity is required for rearrangements~\cite{KeysAbateNP07, AdamPRE07}.  If the dynamical heterogeneities are one-dimensional string-like objects, as is the usual case, then the growth of $n^*$ can be thought of as proportional to a growing dynamical correlation length.

To investigate the possibility of the growth of fluctuating dynamics as a function of depth along the sidewall for the steady heap flow of previous sections, we employ the above formalism using a novel image-based overlap order parameter.  Recall that the raw image slices obtained earlier for computation of $g(x)$ and $v_x$ have grayscale variations indicative of grain positions.  It is not necessary to identify and track the particle coordinates.  Instead, much more simply, a suitable overlap order parameter may be defined by forming the temporal autocorrelation of a sequence of image slices, subtracting the baseline, and dividing by the intercept:
\begin{equation}
Q_t(\tau) = { \langle I_i(t+\tau)I_i(t) \rangle -  \langle I_i(t) \rangle^2
                       \over
            	    \langle I_i(t)^2 \rangle -  \langle I_i(t) \rangle^2     }
\label{oop}
\end{equation}
where the average is over all pixel indices $i$ in the image slice.  By construction this decays from one to zero over a timescale $r/v_x$ set by grain radius divided by average flow speed.  More generally the length scale probed by the ``order parameter'' $Q_t(\tau)$ is given by the correlation length of intensity fluctuations in the grayscale images, which here is about one grain radius.  The actual decay vs $\tau$ for a given start time $t$ varies according to the number fluctuations and the size of the heterogeneities, as reflected in the normalized variance $\chi_4(\tau)$.  Here the image slices are 22 pixels thick and have about $N=150$ grains.

The experimental results for $Q(\tau) = \langle Q_t(\tau) \rangle$ are displayed on linear-logarithmic axes in Fig.~\ref{QX4_t}(a) for a variety of depths $z$.  In all cases $Q(\tau)$ decays monotonically from one but then exhibits attenuated oscillations (not displayed) around zero, similar to the behavior of the pair correlation function.  Nearer the top free surface for faster flow, the decay of $Q(\tau)$ is more rapid; deeper down the decay becomes progressively slower, as expected.  Except for the final oscillations around zero, the decay shape may be accurately fit to a compressed exponential, $Q(\tau)=\exp[-(\tau/\tau_0)^\beta]$, as shown by the dashed curves.  Fitting results for the exponent $\beta$ and the decay time $\tau_0$ are displayed vs depth in Figs.~\ref{btn_z}(a,b).   Near the top surface, the exponent is slightly less than $\beta=2$; such a value would be expected for purely convective flow, as verified by the spatial intensity autocorrelation of individual video image frames.  The exponent then decreases with depth and saturates at a value near $\beta=1$; such a value would be expected if there were considerable jostling in addition to flow, as when diffusion dominates convection.  The fitted decay times $\tau_0$ increase monotonically with depth by many orders of magnitude.  As shown in Fig.~\ref{btn_z}(b) they closely track the time scales $r/v_x$ and $1/\dot\gamma$, but are smaller by respective factors of about 2 and 20.

The corresponding results for $\chi_4(\tau)$ vs $\tau$ are displayed on logarithmic axes in Fig.~\ref{QX4_t}(b), underneath $Q(\tau)$, and are identically color-coded according to depth.  These susceptibilities all begin at zero for short delays, rise to a peak ${\chi_4}^*$ at a delay time $\tau^*$ that coincides closely in Fig.~\ref{btn_z}(b) with the fitted decay time $\tau_0$, and then fall back to zero.  Most significantly, the peaks in $\chi_4(\tau)$ increase in height deeper into the sample where the relaxation time is slower.   Therefore, the dynamics do indeed become more unsteady and spatially heterogeneous on approach to jamming far below the top surface.

The size $n^*$ of heterogeneities may now be deduced at each depth from the peak value ${\chi_4}^*$ using Eq.~(\ref{nstar}).   For this we estimate the values of $Q_0$ and $Q_1$ from averages of $Q_i(\tau^*)$ computed separately for pixels $i$ with $Q_i(\tau^*)<Q^*$ and with $Q_i(\tau^*)>Q^*$, respectively.  This gives $Q_0=0.2\pm0.1$ and $Q_1=0.7\pm0.1$, where the uncertainties represent the uncorrelated spread of values for three different depths; the total average is $Q^*=0.4\pm0.1$.  With only slight adjustment of these averages within the prescribed uncertainties, the resulting size of heterogeneities at the top free surface is $n^*=1$, as expected far from jamming.   All results for $n^*$ are plotted vs depth in Fig.~\ref{btn_z}(c) on semi-logarithmic axes.    The data increase with depth by about two orders of magnitude over the range of depths measured, appearing to grow without bound.  The functional form of this divergence is nearly exponential, $n^*\sim\exp[z/z_n]$ where $z_n=0.17$~cm, as illustrated by the solid line.  As a word of caution, note that the value of $n^*$ becomes larger than the width of the constant-depth image strips; therefore the true heterogeneity sizes may be larger than our estimates.

\begin{figure}
\includegraphics[width=3.00in]{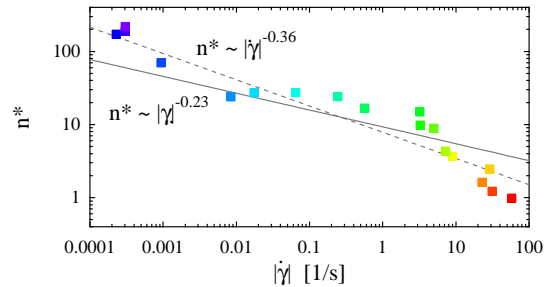}
\caption{(Color online). The number $n^*$ of grains in a dynamical heterogeneity plotted vs strain rate.   Each point corresponds to a different depth, as indicated by the same color code as in the previous figures.  The solid gray line represents a fit to $n^*\propto |\dot\gamma|^{-0.23}$ with exponent fixed according to the simulation result of Hatano~\protect{\cite{Hatano0804}}.  The solid dashed line represent the best power-law fit.}
\label{n_g}
\end{figure}

Recently Hatano~\cite{Hatano0804} simulated the rheology and microscopic dynamics of a three-dimensional granular material under steady shear.  From the instantaneous velocity field, he finds a correlation length that diverges as a power law on approach to jamming at zero strainrate: $\xi \sim |\dot\gamma|^{-0.23}$.  For string-like dynamical heterogeneities, the value of $n^*$ scales as $\xi$.  Therefore to compare directly with Hatano's results we plot $n^*$ vs strainrate in Fig.~\ref{n_g} on logarithmic axes.  The results are reasonably well-described by Hatano's power-law at intermediate strain rates, as shown by the line with slope $-0.23$; a somewhat better fit over the whole range is obtained with exponent $-0.36$.   However the meaning of direct comparison is not clear since the conditions are not identical:  In simulation the shear is uniform and the pressure increases with strainrate as $P\sim\dot\gamma^{1/2}$, whereas in experiment the shear is not uniform and the pressure increases with depth and hence decreasing strainrate.  In the next section we discuss another comparison method.

Before proceeding, we note that while the results for $n^*$ vs depth in Fig.~\ref{btn_z}(c) are clearly monotonic and well-fit to an exponential for small $z$, it appears that one oscillation cycle around the exponential fit develops at the deepest depths examined.  Non-monotonic behavior for the peak in $\chi_4^*$ has been reported for granular \cite{DauchotEPL08a, DauchotEPL08b} and colloidal \cite{BallestaNP08} systems at high packing fractions where, in both cases, some particles do not rearrange or are highly constrained.   Here, data are taken only where the grains rearrange.  And structurally, there is no difference in the packing configuration vs depth as evident by eye or in the measures in Figs.~\ref{akg}(b,c).  Therefore we suppose that the oscillation is an artifact of poor sampling statistics due to both the slow flow and the large value of $n^*$.  Specifically at the greatest depths, the speed is roughly one grain radius per 1000~s, the strain rate is roughly $10^{-4}$/s, and the value of $n^*$ is roughly 100.  According to these orders of magnitude, in a 24~hour run, the grains move only 100 radii and accumulate a strain of only 10.


\section{Discussion}

\begin{figure}
\includegraphics[width=3.00in]{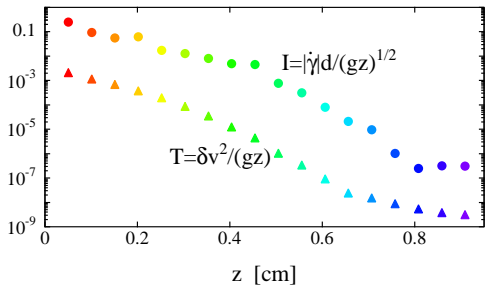}
\caption{(Color online).  Dimensionless effective temperature $T$ and inertial number $I$ plotted vs depth.}
\label{TI_z}
\end{figure}

\begin{figure}
\includegraphics[width=3.00in]{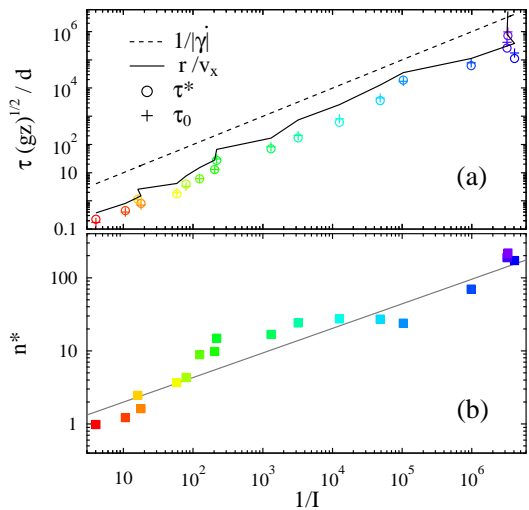}
\caption{(Color online).  The dependence on reciprocal inertia number $1/I=\sqrt{gz}/(\dot \gamma d)$ of (a) the dynamical times of Fig.~\protect{\ref{btn_z}}(b), divided by the characteristic time $d/\sqrt{gz}$, and (b) the heterogeneity domain size $n^*$.  The solid line in (b) is a power-law fit, $n^* \propto (1/I)^{0.33\pm0.02}$.}
\label{tn_I}
\end{figure}

We have measured two quantities that describe the microscopic dynamics of granular heap flow and how it changes with depth: the relaxation times $\tau^* \approx \tau_0 \approx 0.5 r/v_x \approx 0.05/\dot\gamma$, and the number $n^*$ of grains in a dynamical heterogeneity.  On approach to jamming, all of these quantities grow without apparent bound.  To investigate the universality of this jamming transition in comparison with other systems, we now seek the dimensionless parameters that control these divergences at increasing depth.   In general structural rearrangements can arise due to stochastic ``thermal'' fluctuations and also due to shear.  The jamming hypothesis is that physical quantities ought to be universal functions of the dimensionless (effective) temperature $T=\rho \delta v^2/P$ and inertia number $I=\dot\gamma d/\sqrt{P/\rho}$.  Here the pressure increases with depth as $P=\rho g z$,  since the maximum depth is comparable to the plate spacing, so these reduce to
\begin{eqnarray}
  T &=& \delta v^2/(gz), \label{T}\\
  I  &=& \dot\gamma d/\sqrt{g z}.\label{I}
\end{eqnarray}
The latter implies that the relevant characteristic time scale is set by pressure and grain size and density as $\tau_c = d/\sqrt{gz}$, which decreases with depth.  The two candidate control parameters are evaluated using experimental results from previous sections, and are plotted vs depth in Fig.~\ref{TI_z}.  Both are largest near the surface, and decrease with depth.  Note that $T$ is much less than one, and furthermore is a factor of 100 or more times smaller than $I$ at all depths.  Therefore we conclude that $T$ is irrelevant and that the behavior is controlled essentially solely by $I$.  This makes sense because the stochastic grain motion is fueled, ultimately, by the gravity-driven shear flow.

The microscopic relaxation times, rendered dimensionless by comparison with the characteristic time $\tau_c = d/\sqrt{gz}$, and the dimensionless size scale $n^*$ of the dynamical heterogeneities, are both plotted vs inverse inertial number in Figs.~\ref{tn_I}(a,b).   This is identical in spirit to the Arrhenius plots constructed in Fig.~4 of Ref.~\cite{AbatePRL08} for a monolayer of gas-fluidized spheres.   For the time scales in Fig.~\ref{tn_I}(a), note that the dashed line representing $1/\dot\gamma$ scaled by $\tau_c$ is identical to $1/I$.  Since $r/v_x$, $\tau^*$, and $\tau_0$ are proportional to $1/\dot\gamma$,  all the dimensionless microscopic time scales therefore grow linearly with the inverse inertial number.   This makes sense because $I$ is the ratio of inertial to confining stresses, which reduces to a ratio of time scales. 

The heterogeneity length scale in Fig.~\ref{tn_I}(b) also appears to grows as a power of inverse inertia number, but with a systematic oscillation as noted earlier in the discussion of the $n^*$ vs depth plot in Fig.~\ref{btn_z}(c).  The best fit is $n^*\propto (1/I)^{0.33\pm0.02}$, as shown by the solid line.  In remarkable agreement, Hatano's simulation result \cite{Hatano0804} $P\sim\dot\gamma^{1/2}$ gives the inertial number as $I \propto \dot\gamma/\sqrt{P} \sim \dot\gamma^{3/4}$; therefore his result $\xi\sim\dot\gamma^{-1/4}$  can be expressed as $\xi\sim(1/I)^{1/3}$.  Furthermore, Staron et al. have recently found that the length scale of velocity correlations behaves as $\xi\sim(1/I)^{0.3}$ for gravity-driven granular flow down an incline \cite{Staron}.


\section{Conclusion}

In this manuscript we have reported on several significant advances.  In terms of experimental technique, we made the first application of Speckle-Visibility Spectroscopy to heap flow.  Furthermore, perhaps more importantly, we introduced a new method for the measurement of dynamical heterogeneities based on video image correlations.  It obviates the need for particle-tracking, and could be used for any system in which the packing units may be at least partially resolved in images.   In terms of concept, we recognized that the depth below the surface of a freely-flowing granular heap may be considered as a qualitatively new experimental control parameter for the approach to jamming.  This complements prior work in which jamming is approached by variation of temperature or density.  In terms of discovery, we observed that microscopic velocity fluctuations  grow relative to the mean hydrodynamic flow profile as both slow down toward jamming.   We also discovered that dynamical heterogeneities exist and have size that increases as a function of depth.  Both such behaviors have been noted before for systems with uniform driving, but not previously for heap flow.  The velocity fluctuations and dynamical heterogeneities ought to result in distinct but measurable local pressure fluctuations, which would be interesting to compare with those seen in compressed granular media under shear \cite{BehringerNature05, CorwinPRE08}.   Dynamical heterogeneities and their growth with depth for heap flow are thus a crucial rate-dependent microscopic feature, incorrectly neglected in continuum granular hydrodynamic approaches.  It may be an unrecognized reason for overwhelming focus on flow down a rough inclined plane, which is not a more natural or important situation than heaps but which allows the imposition of a no-flow boundary condition.  Our results  for microscopic dynamics are further important in terms of similarity with other systems when plotted vs a dimensionless physical control parameter.  This significantly expands and solidifies the universality of the jamming concept.


We thank K. Kishinawa for experimental assistance (2007), and A.J. Liu for helpful discussions (2009). This work was supported by the JSPS Postdoctoral Fellowships for Research Abroad (H.K. 2005-2007) and the NSF through Grant Nos. DMR-0514705 and DMR-0704147 (D.J.D.).

\bibliography{JamDepthRefs}

\end{document}